\documentclass[12pt, a4paper]{article}
\usepackage{epsf}
\usepackage{cite}
\usepackage{amsmath,amssymb}
\input{colordvi.tex}
\usepackage[usenames,dvipsnames]{color}
\usepackage[dvips]{graphicx}
\usepackage{comment}
\begin{document}

\newcommand{\lsim}{\stackrel{<}{_\sim}}
\newcommand{\gsim}{\stackrel{>}{_\sim}}
\newcommand{\rem}[1]{{ {\color{red} [[$\spadesuit$ \bf #1 $\spadesuit$]]} }}

\renewcommand{\theequation}{\thesection.\arabic{equation}}
\renewcommand{\thefootnote}{\fnsymbol{footnote}}
\setcounter{footnote}{0}


\def\thefootnote{\fnsymbol{footnote}}
\def\a{\alpha}
\def\b{\beta}
\def\c{\varepsilon}
\def\d{\delta}
\def\e{\epsilon}
\def\f{\phi}
\def\g{\gamma}
\def\h{\theta}
\def\k{\kappa}
\def\l{\lambda}
\def\m{\mu}
\def\n{\nu}
\def\p{\psi}
\def\q{\partial}
\def\r{\rho}
\def\s{\sigma}
\def\t{\tau}
\def\u{\upsilon}
\def\v{\varphi}
\def\w{\omega}
\def\x{\xi}
\def\y{\eta}
\def\z{\zeta}
\def\D{\Delta}
\def\G{\Gamma}
\def\H{\Theta}
\def\L{\Lambda}
\def\F{\Phi}
\def\P{\Psi}
\def\S{\Sigma}
\def\me{\mathrm e}

\def\o{\over}
\def\beq{\begin{eqnarray}}
\def\eeq{\end{eqnarray}}
\newcommand{\vev}[1]{ \left\langle {#1} \right\rangle }
\newcommand{\bra}[1]{ \langle {#1} | }
\newcommand{\ket}[1]{ | {#1} \rangle }
\newcommand{\bs}[1]{ {\boldsymbol {#1}} }
\newcommand{\mc}[1]{ {\mathcal {#1}} }
\newcommand{\mb}[1]{ {\mathbb {#1}} }
\newcommand{\EV}{ {\rm eV} }
\newcommand{\KEV}{ {\rm keV} }
\newcommand{\MEV}{ {\rm MeV} }
\newcommand{\GEV}{ {\rm GeV} }
\newcommand{\TEV}{ {\rm TeV} }
\def\diag{\mathop{\rm diag}\nolimits}
\def\Spin{\mathop{\rm Spin}}
\def\SO{\mathop{\rm SO}}
\def\O{\mathop{\rm O}}
\def\SU{\mathop{\rm SU}}
\def\U{\mathop{\rm U}}
\def\Sp{\mathop{\rm Sp}}
\def\SL{\mathop{\rm SL}}
\def\tr{\mathop{\rm tr}}
\def\sp{\;\;}

\def\IJMP{Int.~J.~Mod.~Phys. }
\def\MPL{Mod.~Phys.~Lett. }
\def\NP{Nucl.~Phys. }
\def\PL{Phys.~Lett. }
\def\PR{Phys.~Rev. }
\def\PRL{Phys.~Rev.~Lett. }
\def\PTP{Prog.~Theor.~Phys. }
\def\ZP{Z.~Phys. }

\def\lrf#1#2{ \left(\frac{#1}{#2}\right)}
\def\lrfp#1#2#3{ \left(\frac{#1}{#2} \right)^{#3}}

\newcommand{\lmk}{\left(}  
\newcommand{\rmk}{\right)}
\newcommand{\lkk}{\left[}  
\newcommand{\rkk}{\right]}
\newcommand{\lhk}{\left \{ }  
\newcommand{\rhk}{\right \} }
\newcommand{\del}{\partial}  
\newcommand{\la}{\left\langle} 
\newcommand{\ra}{\right\rangle}
\newcommand{\dd}{\text{d}}


\begin{titlepage}

\begin{center}

\hfill TU-1021, IPMU16-0058\\

\vskip .75in

{\Large \bf 
Diphoton excess from hidden U(1) gauge symmetry with large kinetic mixing
}

\vskip .75in

{\large  Fuminobu Takahashi$^{a,b}$, Masaki Yamada$^{a}$ and Norimi Yokozaki$^{a}$}

\vskip 0.25in

\begin{tabular}{ll}
$^{a}$ &\!\! {\em Department of Physics, Tohoku University,}\\
&{\em Sendai, Miyagi 980-8578, Japan}\\[.3em]
$^{b}$ &\!\! {\em Kavli IPMU (WPI), UTIAS,}\\
&{\em The University of Tokyo,  Kashiwa, Chiba 277-8583, Japan}\\[.3em]
\end{tabular}

\end{center}
\vskip .5in

\begin{abstract}
We show that the $750$\,GeV diphoton excess can be explained by introducing vector-like
quarks and hidden fermions charged under a hidden U(1) gauge symmetry, which has a relatively
large coupling constant as well as  a significant kinetic mixing with U(1)$_Y$.
With the large kinetic mixing, 
the standard model gauge couplings unify around $10^{17}$\,GeV, suggesting the grand unified theory without too rapid proton decay.
Our scenario predicts events with a photon and missing transverse momentum, and 
its cross section is related to that for the diphoton excess through the kinetic mixing. 
We also discuss other possible collider signatures and cosmology,
including various ways to evade constraints on exotic stable charged particles.
In some cases where the $750$\,GeV diphoton excess is due to diaxion decays,
our scenario also predicts triphoton and tetraphoton signals.
\end{abstract}

\end{titlepage}


\renewcommand{\thepage}{\arabic{page}}
\setcounter{page}{1}
\renewcommand{\thefootnote}{\#\arabic{footnote}}
\setcounter{footnote}{0}


\section{Introduction}
\setcounter{equation}{0}

The diphoton excess with an invariant mass around $750$\,GeV was recently reported
 by the ATLAS~\cite{ATLAS} and CMS~\cite{CMS:2015dxe}  collaborations;
for a spin-$0$ particle with a narrow width approximation,  the local significance
is estimated to be $3.9 \sigma$ and $2.9\sigma$, respectively.
 While more data is certainly needed to confirm if the signal
is real or just a statistical fluke, its high statistical significance in the clean analysis using photons
triggered enthusiasm and exuberance for the new physics beyond the standard model (SM), followed 
by the appearance of many theoretical papers. 

Among various models proposed so far, the simplest one is to include a gauge singlet (pseudo)scalar coupled to 
vector-like quarks and/or leptons (see e.g. Refs.~\cite{Harigaya:2015ezk,Buttazzo:2015txu,Franceschini:2015kwy} for the early works).\footnote{
It is still a puzzle why such a (pseudo)scalar coupled to gluons and photons exists in nature. One possible
answer is to relate it to the QCD axion (or its bosonic partner, saxion) which solves the strong CP problem~\cite{Higaki:2015jag,Chiang:2016eav,
Higaki:2016yqk,Gherghetta:2016fhp}.
}
In this model, the (pseudo)scalar is  produced via gluon fusion and decays into a pair of photons through
one-loop diagrams with the extra quarks/leptons running in the loop.  The diphoton excess can be explained if the product 
of the production cross section times branching ratio to two photons is in the range of $5-10$ fb.
This gives a preference to a relatively large branching fraction to diphotons, which necessitates either 
multiple extra matter fields and/or large hypercharges $(Y \gtrsim 1)$ of the extra field running in the loop.

In this paper we consider a possibility that the large hypercharges are originated from 
unbroken hidden U(1)$_H$ gauge symmetry which has a relatively strong gauge coupling and
a significant kinetic mixing with U(1)$_Y$. Then, hidden fermions acquire large hypercharges
due to the kinetic mixing, and the induced hypercharges are generically irrational. We will show that the diphoton
excess can be explained by the (pseudo)scalar coupled to gluons and photons though the extra quark/hidden
fermion loop diagrams. 

Our scenario is based on a rather simple U(1)$_H$ extension of the standard model, which enables
us to make a definite prediction that can be tested soon at the LHC Run-2. 
Since the hidden fermions are charged under U(1)$_H$, the (pseudo)scalar responsible for the diphoton excess
can also decay into $\gamma \gamma'$, where $\gamma'$ denotes the hidden photon. Thus,
our scenario predicts events with a photon and missing momentum,\footnote{
See Ref.~\cite{Tsai:2016lfg} for a related work.
} and we will see that its production cross 
section times branching fraction is simply related to that for the diphoton excess through the kinetic mixing.
The events with a photon and missing momentum have been searched for at the LHC 
Run-1~\cite{Aad:2014tda,Khachatryan:2014rwa} 
and Run-2~\cite{Aaboud:2016uro}, and there
is an upper bound on the production cross section. We will see that the experimental bound places 
a {\it lower} bound on the kinetic mixing.

A large kinetic mixing with U(1)$_H$ is known to modify  the normalization of the hypercharge so that
the gauge coupling unification is improved~\cite{Redondo:2008zf}.
We will show that this is indeed the case in our model, taking account of contributions of 
the extra matter fields to the renormalization group (RG) equations.\footnote{
In supersymmetric models with the grand unification, 
the diphoton excess may indicate the light gluino of 2\,-\,3\,TeV, which originates from changes of RG equations with extra matter fields~\cite{diphoton_susy_gut}.
}
The hidden fermions
acquire hypercharges through the kinetic mixing, and they are cosmologically stable. 
 Such stable exotic charged particles, if produced abundantly in the early Universe, could affect the big 
 bang nucleosynthesis (BBN)~\cite{Pospelov:2006sc,Kohri:2006cn,Pospelov:2008ta} as well as cosmic 
 microwave back ground radiation (CMB)~\cite{Dubovsky:2003yn,Dolgov:2013una}. 
 Also there are various experimental searches for exotic fractional or multi charged particles~\cite{Burdin:2014xma,atlas_stable,cms_stable}. 
 We will discuss several possibilities to evade those constraint. Finally, we will discuss other possible ``diphoton'' excesses
at different energies if the $750$\,GeV diphoton excess is due to diaxion decays.

The rest of this paper is organized as follows. In Sec.~2 we show that the diphoton excess
can be explained by introducing a gauge singlet (pseudo)scalar and vector-like quarks
and hidden fermions, the latter of which is charged under U(1)$_H$. In Sec.~3 we study
the gauge coupling unification in the presence of the large kinetic mixing.
Cosmological implications are discussed in Sec.~4. The last section is devoted for
 conclusions.

\section{Kinetic mixing with hidden U(1)}
\setcounter{equation}{0}
\subsection{Preliminaries}
Let us first quickly review the effect of a kinetic mixing between two U(1)s, U(1)$_1$ and U(1)$_2$.
We will shortly apply the results to the kinetic mixing between U(1)$_Y$ and a hidden U(1) gauge
symmetry, U(1)$_H$.

Let us consider the Lagrangian~\cite{Holdom:1985ag},
\beq
{\cal L} = - \frac{1}{4} F'^{\mu \nu}_1 F'_{1\mu\nu} -  \frac{1}{4} F'^{\mu \nu}_2 F'_{2\mu\nu} 
- \frac{\chi}{2} F'^{\mu \nu}_1 F'_{2\mu\nu},
\eeq
where $F'^{\mu \nu}_i \equiv \partial^\mu A'^{\nu}_{i} - \partial^\nu A'^{ \mu}_{i}~(i=1,2)$ represent the field 
strength of U(1)$_i$, and $\chi$ is the kinetic mixing between
them. The kinetic mixing can be removed by the following transformation,
\beq
{A_1^\mu}' &=& \frac{{A_1^\mu}}{\sqrt{1-\chi^2}}, \\
{A_2^\mu}' &=& {A_2^\mu} - \frac{\chi}{\sqrt{1-\chi^2}} {A_1^\mu},
\eeq
where ${A_1^\mu}$ and ${A_2^\mu}$ are canonically normalized gauge fields. 
Hereafter we call this new basis ($A_i^\mu$) 
as the canonical basis to distinguish it from the original basis ($A'^\mu_i$). 
In the canonical basis,  gauge couplings $e_1$ and $e_2$ are written in terms of 
the kinetic mixing $\chi$ and the
gauge couplings in the original basis $e'_1$ and $e'_2$ such as 
\beq
{e_1} &=& \frac{e_1'}{\sqrt{1-\chi^2}}, 
\label{e'}
 \\
{e_2} &=& {e_2'}.
\eeq
In the canonical basis, any matter fields charged under U(1)$_1$ in the original basis are still coupled to $A_1^\mu$
with a rescaled gauge coupling, $e_1$. On the other hand, the matter field charged under U(1)$_2$ acquires
an induced charge of U(1)$_1$ in the canonical basis. For instance, 
\beq
q_2 e_2' \bar \psi \gamma^\mu \psi A_{2\mu}' 
= q_2 e_2 \bar \psi \gamma^\mu \psi {A_{2\mu}} 
- \frac{\chi}{\sqrt{1-\chi^2}} q_2 e_2 
\bar \psi \gamma^\mu \psi {A_{1\mu}}.
\eeq
Thus, through the kinetic mixing,  a matter field with a charge $q_2$ under U(1)$_2$ acquires a charge, $- \frac{\chi}{\sqrt{1-\chi^2}} \frac{q_2 e_2}{e_1}$,
under U(1)$_1$ in the canonical basis. The induced charge is generically irrational, and can be larger than unity depending on 
the relative size of the gauge couplings and the kinetic mixing.

\subsection{Diphoton excess}
Now we apply the above result to SU(2)$_L \times$ U(1)$_Y \times$ U(1)$_H$, where
we assume a kinetic mixing $\chi$ between U(1)$_Y$ and U(1)$_H$.  Suppose that there is a hidden
matter field $\psi$ with a charge $q_H$ under U(1)$_H$. Then, in the canonical basis, the
hidden matter field acquires an electric charge,
\beq
q_{\rm eff} = - \frac{\chi}{\sqrt{1-\chi^2}} \frac{q_H e_H}{e_{\rm EM}} \cos\theta_w,
\label{q_eff}
\eeq
where $e_{\rm EM}$ and $e_H$ are gauge couplings of U(1)$_{\rm EM}$ and U(1)$_H$, respectively, and $\theta_w$
represents the weak mixing angle, $\sin^2 \theta_w \simeq 0.23$. The induced electric charge is generally
irrational, and it can be relatively large if the hidden U(1)$_H$ is more strongly coupled than the electromagnetic
coupling, i.e., $e_H > e_{\rm EM}$ and if the kinetic mixing is large,  $\chi = {\cal O}(0.1)$.

To be concrete, let us consider a variant of the volksmodel, where a complex scalar $\Phi$ is coupled to
$n_q$ vector-like extra quarks $(D, \bar D)$ and $n_\psi$ hidden fermions $(\psi, \bar \psi)$;
\beq
- {\cal L} = y_q \Phi \sum_{i=1}^{n_q} \bar{D}_i D_i + y_\psi \Phi \sum_{i=1}^{n_\psi} \bar \psi_i \psi_i + h.c.,
\eeq
where the subscript $i$ denotes flavor of the extra quarks and hidden fermions. 
The charge assignment of these extra matter fields is given in Table \ref{tab:charge}.\footnote{
One may impose an approximate global U(1) symmetry to ensure the above interaction~\cite{Higaki:2015jag}.}  
Here, 
we focus on the case in which $\psi$ and $\bar{\psi}$ are SM gauge singlets in the original basis 
and have charges $q_H$ and $-q_H$ under U(1)$_H$, respectively. 
Then the hypercharge of $\psi_i$ is induced solely by the kinetic mixing 
as in Eq.~(\ref{q_eff}). 

\begin{table}[tb]
\begin{center} {\tabcolsep = 2mm
	\begin{tabular}{c|cccc} \hline
		\rule[0mm]{0mm}{4.0mm} &  $D_i$ & $\bar{D}_i$ & $\psi_i$ & $\bar{\psi}_i$  \\ \hline
		SU(3) & ${\bf 3}$ & ${\bf \bar{3}}$  & {\bf 1} & {\bf 1} \\
		U(1)$_Y$ & $a$ & $-a$  & $0 (q_{\rm eff})$ & $0 (- q_{\rm eff})$  \\
		U(1)$_H$ & 0 & 0  &$q_H$ & $-q_H$  
		\\ \hline
	\end{tabular} }
\end{center}
\caption{Charge assignment of the extra fermions in the original (canonical) basis.}
\label{tab:charge}
\end{table}

We assume that $\Phi$ develops a non-zero expectation value in the vacuum,
\beq
\Phi = \frac{f + s}{\sqrt{2}} e^{i \phi/f}, 
\eeq
where $s$ and $\phi$ denote the radial and phase degrees of freedom, respectively, and
$f$ is the decay constant. Then, the extra matter fields have masses of $y_{q, \psi} f/\sqrt{2}$. 
While the diphoton excess can be explained by either $s$ or $\phi$,
we will focus on $\phi$ in the following analysis. Our results can be straightforwardly applied
to $s$ except for a possibly large branching fraction of $s$ decaying into a pair of $\phi$.\footnote{
The decay of $s$ into a pair of $\phi$ can be suppressed by introducing $\Phi_1$ and $\Phi_2$ with opposite
PQ charges, if they respect an approximate $Z_2$ exchange symmetry, $\Phi_1 \leftrightarrow \Phi_2$.
} 
Hereafter, we assume the mass of $\phi$, denoted as $m_\phi$, to be $750 \ \GEV$.

The field $\phi$ can decay to gluons and photons via 1-loop diagram and 
their decay rates are given by 
\begin{eqnarray}
&&\Gamma( \phi \to gg) = 
8 \lmk \frac{\alpha_3}{8\pi f} \rmk^2 \frac{ m_{\phi}^3}{\pi}
\left| \sum_i  \frac{1}{4} A_{1/2}(x_i) \right|^2, 
\\
&&\Gamma( \phi \to \gamma \gamma) = 
\lmk \frac{\alpha_{\rm EM}}{8\pi f} \rmk^2 \frac{ m_{\phi}^3}{\pi}
\left| \sum_i  \frac{Q_{i}^2}{2} A_{1/2}(x_i) \right|^2, 
\end{eqnarray}
where $\alpha_{\rm EM}$ and $\alpha_3$ are the electroweak and strong gauge coupling strength, respectively, 
$Q_i$ is an electric charge of the $i$-th particle in the loop, and $x_i \equiv 4 m_i^2 / m_\phi^2$. 
The form factor $A_{1/2}$ is given by 
\begin{eqnarray}
A_{1/2} (x) &=&  2x  \arcsin^2 \lmk 1/\sqrt{x} \rmk \ \ {\rm for} \ \, x \geq 1, 
\end{eqnarray}
which satisfies $A_{1/2} ( \infty ) = 2$. 
Then, the production cross section for $pp \to \phi \to \gamma \gamma$ is estimated as
\begin{eqnarray}
\sigma(pp\to \phi + X) {\rm Br}(\phi \to \gamma\gamma)  &\simeq& K \cdot \, \frac{\pi^2}{8 m_\phi} \frac{1}{s} \, 
\Gamma( \phi \to gg) {\rm Br}( \phi \to \gamma\gamma) C_{gg}, \nonumber \\
C_{gg} &\equiv& \int_0^1 d x_1 \int_0^1 d x_2 f_g(x_1) f_g (x_2) \delta(x_1 x_2 - m_{\phi}^2/s),
\end{eqnarray}
where $K$ denotes the K-factor, $\sqrt{s}=13$\,TeV and $m_{\phi}=750$\,GeV. 
Taking the factorization scale to be $0.5 m_{\phi}$, $C_{gg} \approx 1904$ using {\tt MSTW2008NNLO}~\cite{mstw2008}.
Since $\Gamma(\phi \to gg)$ is much larger than  $\Gamma(\phi \to \gamma \gamma)$ in the parameter region of our interest, we use an approximation, $\Gamma( \phi \to gg) {\rm Br}( \phi \to \gamma\gamma) \approx \Gamma( \phi \to \gamma\gamma)$, resulting in 
\begin{eqnarray}
\sigma(pp\to \phi + X) {\rm Br}(\phi \to \gamma\gamma) 
 \approx K \cdot 7.2\, {\rm fb} \, 
\left( \frac{\Gamma(\phi \to \gamma\gamma)}{10^{-3}\, {\rm GeV}} \right).
\end{eqnarray}
In a simple case of $a=0$, the cross section is calculated as 
\beq
\sigma(pp\to \phi + X) {\rm Br}(\phi \to \gamma\gamma) 
\simeq 3.5 \,  {\rm fb} \, \lrfp{f}{800{\rm\,GeV}}{-2} \left(\frac{K}{1.5}\right) 
\lmk \frac{q_{\rm eff}^2 n_\psi}{4} \rmk^2, 
\eeq
where we assume $A_{1/2} (x_\psi) \simeq 2$ and the K-factor is estimated as $K \approx 1.5$ (see e.g. \cite{Franceschini:2015kwy}).
In the case of $a \ne 0$, $q_{\rm eff}^2 n_\psi$ 
should be replaced with $ 3 a^2 n_q +q_{\rm eff}^2 n_\psi$ in the above equation.

In our scenario the pseudoscalar $\phi$ decays into other channels. 
The ratios of the decay of $\phi$ into $\gamma \gamma$, $\gamma Z$, $ZZ$, $\gamma \gamma'$, $Z \gamma'$,
and $\gamma' \gamma'$ are given by
\beq
&&\gamma \gamma : \gamma Z : ZZ : \gamma \gamma' : Z \gamma': \gamma' \gamma' \nonumber \\
&&\simeq 1:2 \tan^2 \theta_w:\tan^4\theta_w:2\frac{\alpha_H}{\alpha_{\rm EM}}
 \left(\frac{k_{\rm mix}}{k}\right)^2: 2\frac{\alpha_H}{\alpha_{\rm EM}}\left(\frac{k_{\rm mix}}{k}\right)^2 \tan^2 \theta_w: \left(\frac{\alpha_H k_H}{\alpha_{\rm EM} k}\right)^2, \nonumber \\
\eeq
with 
\beq
k &=& 3 a^2 n_q + q_{\rm eff}^2 n_\psi, \\
k_H &=& q_H^2 n_\psi, \\
k_{\rm mix} &=& q_{\rm eff} q_H n_\psi. 
\eeq
Here we have dropped the phase space factor.
As the simplest realization of our scenario, let us focus on the case of $a=0$. 
Then, the ratios of the branching fraction of $\phi \to \gamma \gamma'$ and $\phi \to Z \gamma'$ to that of $\phi \to \gamma \gamma$ are 
\beq
\frac{{\rm Br(\phi \to \gamma \gamma')}}{{\rm Br(\phi \to \gamma \gamma)}} &=& 2 \frac{1-\chi^2}{\chi^2} \frac{1}{\cos^2 \theta_w}
\simeq 2.6  \frac{1-\chi^2}{\chi^2}, \\
\frac{{\rm Br(\phi \to Z \gamma')}}{{\rm Br(\phi \to \gamma \gamma)}} &=& 2 \frac{1-\chi^2}{\chi^2} \frac{\tan^2 \theta_w}{\cos^2\theta_w} 
\simeq 0.78  \frac{1-\chi^2}{\chi^2}. 
\eeq
One can relate the production cross section for events with a photon and a hidden photon
to that for diphoton events as
\beq
\sigma(pp\to \phi  \to \gamma'\gamma) 
\simeq 10\,{\rm fb}\, \frac{1-\chi^2}{\chi^2}  
\lrf{\sigma(pp\to \phi  \to \gamma\gamma)   }{4 {\rm fb}}.
\eeq
The events with a photon and missing transverse momentum have been searched for at the LHC, and their 
production cross section is constrained to be below $17.8$ fb $(95 \% {\rm CL})$ by the ATLAS experiment 
at $\sqrt{s} = 13$\,TeV~\cite{Aaboud:2016uro}.  This sets a {\it lower} limit on the kinetic mixing as
\beq
\chi \gtrsim 0.6,
\eeq
where we have set $\sigma(pp\to \phi  \to \gamma\gamma) = 4 \,{\rm fb} $. 
The lower bound on $\chi$ is relaxed if $a \ne 0$. 

Using the bound on the kinetic mixing, one can derive a constraint  on $f$
to explain the diphoton excess,
\beq
 f \gtrsim 
 800 \ \GEV
 \lmk \frac{ n_\psi q_H^2 \alpha_H }{0.08} \rmk 
 \lmk \frac{K}{1.5} \rmk^{1/2} 
 \lmk \frac{\sigma (pp \to \phi \to \gamma \gamma )}{3.5 {\rm \ fb}} \rmk^{1/2}. 
\eeq
Thus, one can explain the diphoton excess by introducing a single vector-like hidden lepton charged
under U(1)$_H$ with a kinetic mixing with U(1)$_Y$, while satisfying the experimental bound on events with a
photon and missing transverse momentum.  In particular, neither many vector-like matter fields
nor large hypercharge in the original basis is needed.

Since fractionally charged particles may be produced via Drell-Yan process, 
their charges and masses are constrained by the LHC experiment. 
The ATLAS and CMS collaborations put the upper bound on the mass of stable particle with an electric charge 
of $(1\,\mathchar`-\,2)|e_{\rm EM}|$~\cite{atlas_stable,cms_stable}. One can evade the constraint for $m_{\psi} \gtrsim 600$\,-\,700\,{\rm GeV} with $|q_{\rm eff}|
= 1 \, \mathchar`- \,2$. 
If $\bar D_i$ mixes with a SM quark for $a=1/3$, it decays into e.g. Higgs and a quark, 
avoiding constraints from $R$-hadron searches.\footnote{
If $\bar D_i$ mixes with the bottom quark, the lower bound on its mass is severe as $\sim$700 GeV~\cite{Aad:2015kqa, Khachatryan:2015gza}. 
However, the bound is much weaker in the case that it mixes with a light quark~\cite{Aad:2015tba}.
}

\section{A model with gauge coupling unification}

In this section, we propose a model consistent with the gauge coupling unification. 
The RG flow of gauge coupling constants is modified by the presence of the large kinetic mixing. 
In addition, 
the normalization of U(1)$_Y$ gauge coupling is affected by the large kinetic mixing 
[see Eq.~(\ref{e'})], where we require that 
the SM gauge couplings are unified in the original basis.

We introduce $N_5$ pair of $SU(5)$ complete multiplets as
\begin{eqnarray}
-\mathcal{L} =  y_D \sum_i^{N_5} \Phi D_i \bar D_i + y_L \sum_i^{N_5} \Phi L_i \bar L_i + y_\psi \sum_i^{n_\psi} \Phi \psi_i \bar \psi_i \, + h.c.,
\end{eqnarray}
where $D_i$ and $\bar L_i$ ($\bar D_i$ and $L_i$) consist of $SU(5)$ multiplets transforming ${\bf 5}$ ($\bar{\bf 5})$ representation. The $U(1)_Y$ charges of $D_i$ and $\bar L_i$ are $-1/3$ and $1/2$, respectively, and they are singlet under the U(1)$_H$ gauge group. Here, $\psi_i$ and $\bar \psi_i$ are only charged under U(1)$_H$ as noted in Table 1.

\begin{figure}[!t]
\begin{center}
\includegraphics[width=.40\textwidth, bb=0 0 178 187]{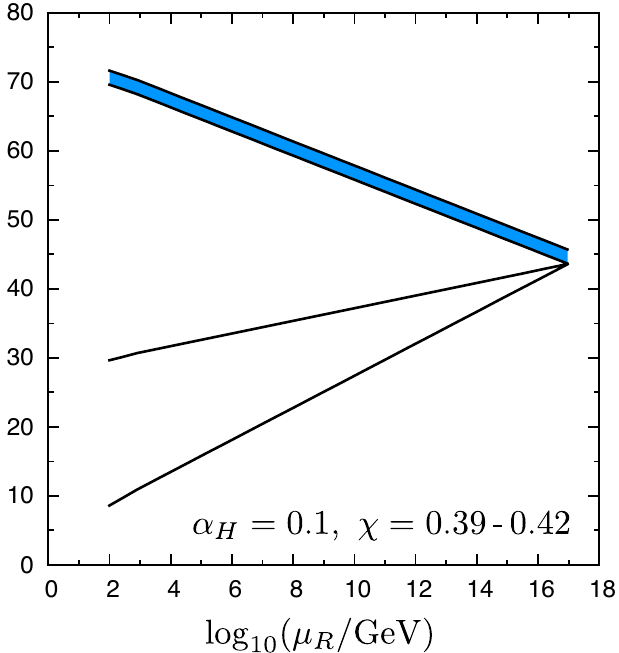}
\caption{
The running of the gauge couplings ($\alpha_1'^{-1},\alpha_2^{-1},\alpha_3^{-1}$) in the original basis
from top to bottom.
Here, $q_H=1$, $n_\psi=1$, $N_5 = 1$, $\alpha_3(m_Z)=0.1185$ and $m_t({\rm pole})=173.34$ GeV.
The masses of the extra matter fields are taken to be $800$\,GeV. We take $\alpha_H=0.1$ and $\chi=0.39$\,-\,$0.42$ at $\mu_R = m_Z$. 
}
\label{fig:gut1}
\end{center}
\end{figure}

\begin{figure}[!t]
\begin{center}
\includegraphics[width=.40\textwidth, bb=0 0 178 187]{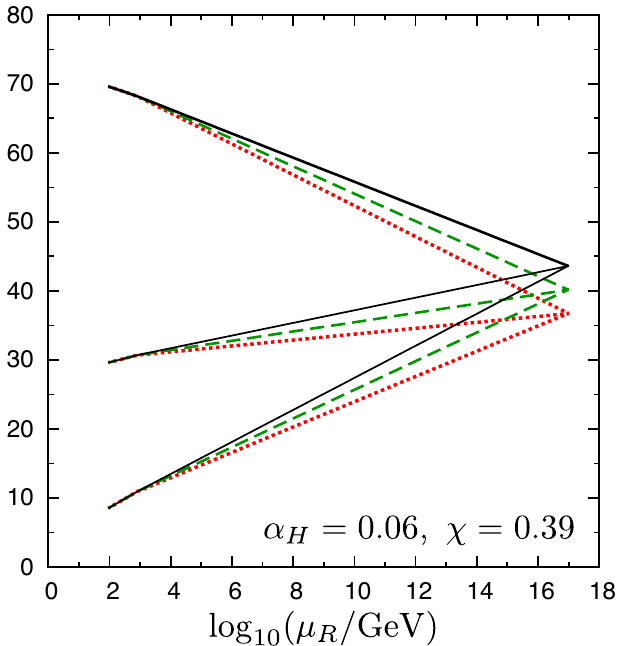}
\caption{
The running of the gauge couplings ($\alpha_1'^{-1},\alpha_2^{-1},\alpha_3^{-1}$) in the original basis
for $N_5=1$, 2 and 3. The solid, dashed and dotted lines show the RG flow of the gauge couplings for $N_5=1, 2$ and 3, respectively.
Here, $\alpha_H=0.06$ and $\chi=0.39$ are taken at $\mu_R = m_Z$. The other parameters are the same as in Fig.~\ref{fig:gut1}. 
}
\label{fig:gut2}
\end{center}
\end{figure}

To calculate the RG  flow, 
it is convenient to write the coupling $q_{\rm eff} g_Y$ as $q_H g_{\rm mix}$. 
Thus, in the canonical basis, a matter field $\Psi$, which collectively denotes $D$, $L$ and $\psi$,  have an interaction with
\begin{eqnarray}
\mathcal{L} = \bar \Psi \gamma^\mu \left[ e_H q_H {A_H}_\mu + (g_Y q_Y  + g_{\rm mix} q_H) {A_Y}_\mu \right] \Psi,
\end{eqnarray}
where $g_{\rm mix}$ is identified with $-e_H \chi/\sqrt{1-\chi^2}$ at the scale where Eq.(2.1) is defined. 
Then, the RG equations are~\cite{Babu:1996vt}
\begin{eqnarray}
\frac{d g_Y}{d t} &=&  \frac{g_Y}{16\pi^2} \left( b_Y g_Y^2  + b_H g_{\rm mix}^2 \right),\nonumber \\
\frac{d e_H}{d t} &=& \frac{e_H}{16\pi^2} \left( b_H e_H^2 \right), \nonumber \\
\frac{d g_{\rm mix}}{d t} &=& \frac{g_{\rm mix}}{16\pi^2} \left( b_Y g_Y^2  + b_H g_{\rm mix}^2 + 2 b_H e_H^2 \right),\end{eqnarray}
with
\begin{eqnarray}
b_H &=& \frac{4}{3} n_\psi  q_H^2 , \nonumber \\ 
b_Y &=&  \frac{41}{6}  + \frac{10}{9} N_5. 
\end{eqnarray}
The coefficients of beta functions for SU(2)$_L$ and SU(3)$_c$ are given by 
\begin{eqnarray}
b_2 &=& -\frac{19}{6} + \frac{2}{3} N_5 , \nonumber \\
b_3 &=& -7 + \frac{2}{3} N_5.
\end{eqnarray}
Here, $t=\log \mu_R$, where $\mu_R$ is a renormalization scale.

We plot the RG flow of couplings $\alpha_3$, $\alpha_2$, and $\alpha'_1$ [$\equiv (1-\chi^2) \alpha_1$] in Fig.~\ref{fig:gut1} and \ref{fig:gut2}, 
where we take the SU(5) normalization of $\alpha_1 \equiv (5/3) \alpha_Y$ 
and assume $q_H = 1$ and $n_\psi = 1$. 
%
%
%
%
In Fig.~\ref{fig:gut1}, we have varied the kinetic mixing slightly, which is represented by the blue band.
We find that the SM gauge couplings are unified at the energy scale of order $10^{17} \ \GEV$, 
which is consistent with the null result of proton decay. The relevant RGEs are given by~\footnote{
Even if there exist fields which have both U(1)$_Y$ and U(1)$_H$ charges, 
the form of the RGE, $\frac{d {\alpha_1'}^{-1}}{dt} = - \frac{b_Y}{2\pi} \left(\frac{3}{5}\right)$, does not change.
}
\begin{eqnarray}
\frac{d {\alpha_1'}^{-1}}{dt} = - \frac{b_Y}{2\pi} \left(\frac{3}{5}\right), \ 
\frac{d \alpha_2^{-1}}{dt} = - \frac{b_2}{2\pi} , \ 
\frac{d \alpha_3^{-1}}{dt} = - \frac{b_3}{2\pi} \, . \label{eq:rge_gut}
\end{eqnarray}
Apparently, the running of $\alpha_1'$ does not depend on $\alpha_H (m_{\psi})$.
In Fig.~\ref{fig:gut2}, we also show the RG flow of the gauge couplings for $N_5=1$, 2 and 3. One can see that the unification point at around $10^{17}$\,GeV is independent of $N_5$.
The hidden gauge coupling $e_H$ remains perturbative up to the GUT scale:
\begin{eqnarray}
\alpha_H(10^{17}\,{\rm GeV}) &=& 
\alpha_H (m_\Psi) \left[1 - \frac{ \alpha_H (m_\Psi)}{2\pi} \ln \frac{10^{17}\,{\rm GeV}}{m_{\Psi}} \right]^{-1} 
\approx 0.21\,(0.09),
\end{eqnarray}
for $\alpha_H(m_\Psi)=0.10\,(0.06)$.

With the extra SU(5) multiplets, the cross section of the diphoton signal 
is modified such as $k \to k + N_5$, $n_q = N_5$, and $a = -1/3$ due to the presence of the GUT multiplet. 
Taking $q_H^2 n_\psi = 1$, $N_5 =1$, $\chi = 0.40$ and $\alpha_H=0.1$, 
the cross section is estimated as 
\begin{eqnarray}
\sigma(pp\to \phi + X) {\rm Br}(\phi \to \gamma\gamma) 
 \approx 3.4 \ {\rm fb} \left( \frac{f}{800 {\rm GeV}}\right)^{-2}. \label{eq:gut_cross}
\end{eqnarray}
Here, the masses for the extra matter fields are taken as $m_{\psi}=m_{D'}=650 \ \GEV$ and $m_L'=450$\,GeV, where $m_{D'}=y_D f/\sqrt{2}$ and $m_{L'}=y_L f/\sqrt{2}$. In the case $N_5=2$, the same cross section of Eq.~(\ref{eq:gut_cross}) is obtained with the smaller $\alpha_H$ and larger $f$: $\alpha_H=0.06$ and $f=980$\,GeV
with the other parameters being the same as the previous case.

The ratio of the branching fraction of $\phi \to \gamma \gamma'$ to that of $\phi \to \gamma \gamma$ 
is suppressed due to the contribution to the latter process from the GUT multiplet:
\begin{eqnarray}
\frac{\sigma(pp \to \phi \to \gamma\gamma')}{\sigma(pp \to \phi \to \gamma\gamma)} \approx \frac{n_\psi^2 q_{\rm eff}^4}{(N_5 \, 4/3   + n_\psi q_{\rm eff}^2)^2}\frac{1-\chi^2}{\chi^2} \frac{2}{\cos^2\theta_w}.
\end{eqnarray}
As a result, 
we find that the present constraint on the events with a photon and missing transverse momentum 
can be suppressed compared to the case without extra SU(5) multiplets. For a set of the parameters consistent with the GUT, $N_5=1$,  $ n_{\psi} q_{\rm eff}^2 =1.92$, $\chi=0.40$~\footnote{
At the GUT scale $10^{17}$\,GeV, $\chi \sim 0.6\,{\mathchar`-}\,0.8$, depending on $e_H$.
} and $\sigma(pp \to \phi \to \gamma\gamma)=3.4$\,fb, we have $\sigma(pp \to \phi \to \gamma\gamma') \approx 16.1$\,fb.  
Obviously, larger $N_5$ leads to a weaker constraint: for $N_5=2$,  $ n_{\psi} q_{\rm eff}^2 =1.15$ and $\chi=0.40$, the relevant cross section is as small as $\sigma(pp \to \phi \to \gamma\gamma') \approx 4.2$\,fb.

Finally, let us comment on a possible generation of the large mixing. 
In SU(5)$_{\rm GUT}\times$U(1)$_H$ model, we may have the following operator: 
\begin{eqnarray}
\frac{1}{M_*}{\rm Tr}(\Sigma_{24} {F^5_{\mu \nu}}) {F}^{\mu \nu}_H, 
\label{origin of mixing}
\end{eqnarray}
where $F^5_{\mu \nu}$ is a gauge field strength of SU(5)$_{\rm GUT}$; $M_*$ is a cut-off scale, and $\Sigma_{24}$ is a GUT breaking Higgs with 
$\left<\Sigma_{24}\right> = {\rm diag}(2,2,2, -3,-3) v_{\rm GUT}$. (Here, $v_{\rm GUT} \sim 10^{17}\,{\rm GeV}$). Therefore, if $M_*$ is somewhat close to $v_{\rm GUT}$, the large mixing between U(1)$_Y$ and U(1)$_H$ can be generated via the above high dimensional operator.
That said, it is fair to admit that obtaining such a large kinetic mixing is highly nontrivial 
in a context of gauge coupling unification. 
This is because we can similarly write down the following operator: 
\begin{eqnarray}
\frac{1}{M_*}{\rm Tr}(\Sigma_{24} {F^5_{\mu \nu}}{F^{5 \mu \nu}}), 
 \label{violating term}
\end{eqnarray}
which could generate a large threshold correction, 
preventing the gauge couplings from precise unification. 
We note however that the 
relative size of these operators Eqs.~(\ref{origin of mixing}) and (\ref{violating term}) depends on detailes of UV phisics.%
\footnote{
For instance, we can consider interactions:
\begin{eqnarray}
\mathcal{L} =  \lambda_5 f_{\bar{\bf 5}}^l \Sigma_{24} f_{\bf 5}^l + M_5 f_{\bar{\bf 5}}^l f_{\bf 5}^l + h.c. \, ,
\end{eqnarray}
where $f_{\bf 5}^l$ has a $U(1)_H$ charge of $q_{H,f}$, and $l=1 \dots N_f$. Then, 
after integrating out $f_{\bf 5}^l$ and $f_{\bar{\bf 5}}^l$, the mixing term is generated as
\begin{eqnarray}
\mathcal{L} \sim \frac{\lambda_5  g_5 e_H q_{H,f} N_f}{16\pi^2 M_{f}}  {\rm Tr}(\Sigma_{24} F^5_{\mu \nu}) F_H^{\mu \nu},
\end{eqnarray}
where $M_f \sim \lambda_5 v_{\rm GUT} + M_5$. 
For $\lambda_5 \sim 4\pi$, $M_f \approx v_{\rm GUT}$, and $(q_{H,f} N_f) \sim 10$, the mixing becomes $\mathcal{O}(1)$.
In addition, Eq.~(\ref{violating term}) is generated as 
\begin{eqnarray}
\mathcal{L} \sim \frac{\lambda_5 g_5^2 N_f}{16\pi^2 M_{f}}  {\rm Tr}(\Sigma_{24} F^5_{\mu \nu} F_5^{\mu \nu}).
\end{eqnarray}
The relative size of these operators depends on $q_{H,f} e_H$, 
so that we can suppress the coupling constant of the latter operators with fixing that of the former operator. 
We thank an anonymous referee for pointing out this issue. 
} 

\section{Cosmology}
\setcounter{equation}{0}

In this section, we explain cosmology of our model. 
The hidden photon decouples from the SM sector at a temperature around 
$m_\psi / 10$ and contributes to the energy density of the Universe as dark radiation, 
whose amount can be measured by future observations of CMB temperature fluctuations. 
Since the fields $\psi_i$ and $\bar{\psi}_i$ have 
fractional charges of U(1)$_Y$, 
the lightest ones are absolutely stable. On the other hand, 
extra quarks (and leptons) mix with SM quarks (leptons), and are not stable as mentioned earlier.   
The abundance of the fractionally charged particles, $\psi_i$ and $\bar{\psi}_i$, 
are severely constrained by various experiments and observations. 
We provide some possibilities to evade these constraints in Sec.~\ref{possibilities}.

\subsection{Dark radiation}

Since the hidden U(1)$_H$ is not broken in our model, 
we predict hidden photon as well as the fractionally charged particles. 
When the temperature is higher than the mass of $\psi$, 
the U(1)$_H$ gauge boson as well as $\psi$ are in thermal equilibrium with the SM plasma. 
Even after the temperature decreases to $m_\psi$, 
the U(1)$_H$ gauge boson may interact with visible photon via photon-photon scatterings. 
Here we quote the low-energy scattering cross section of visible photons: 
\beq
 \frac{\dd \sigma ( \gamma \gamma \to \gamma \gamma) }{\dd \Omega} 
 = \frac{139}{(180 \pi)^2 } \alpha_{\rm EM}^4 \frac{\omega^6}{m_e^8} 
 \lmk 3 + \cos^2 \theta \rmk^2, 
 \label{photon photon}
\eeq
where $m_e$ is the electron mass and 
$\omega$ is the energy of each colliding photon in the frame 
in which the total momentum vanishes and $\theta$ is a scattering angle. 
We expect that 
scatterings between visible photon and hidden photon 
is roughly given by Eq.~(\ref{photon photon}) 
with the replacement of 
$\alpha_{\rm EM}^4 \to q_{\rm eff}^2 \alpha_{\rm EM}^2 \alpha_H^2$ 
and $m_e \to m_\psi$ with an additional $\mathcal{O}(1)$ coefficient. 
We find that this kind of interaction decouples at a temperature 
of order $m_\psi / 10$ for $m_\psi = \mathcal{O}(1) \ \TEV$. 
However, 
the fractionally charged particles $\psi$ may be still in thermal equilibrium with both the SM plasma and 
the hidden sector via Compton scatterings. 
The Compton scattering between $\psi$ and U(1)$_H$ gauge boson 
is decoupled at a temperature satisfying $\sigma_T' n_\psi / H \sim 1$, 
where $\sigma_T'$ is the Thomson scattering rate given by $8 \pi \alpha_H^2 / 3 m_\psi^2$. 
When the number density of $n_\psi$ is determined by the thermal relic density 
[see Eq.~(\ref{annihilation})], 
the combination is rewritten as 
\beq
 \frac{\sigma_T' n_\psi (T) }{ H (T)}
 \simeq \frac{8}{3} \frac{T_f}{T}, 
\eeq
where $T_f$ is the freezeout temperature of $\psi$ ($\approx m_\psi / 25$). 
Thus the U(1)$_H$ decouples from the SM plasma at a temperature of order $m_\psi / 70$.

After the U(1)$_H$ gauge boson decouples from the SM thermal plasma, 
its energy density contributes to the expansion of the Universe as dark radiation. 
Its amount is conventionally expressed by the effective neutrino number $\Delta  N_{\rm eff}$, 
which is calculated as 
\beq
 \Delta N_{\rm eff} = \frac{8}{7} \lmk \frac{g_* (T_D)}{43/4} \rmk^{-4/3}, 
 \label{delta N_eff}
\eeq
where $g_*(T_D)$ is the effective relativistic degrees of freedom at the decoupling temperature $T_D$ 
(see, e.g., Refs.~\cite{Nakayama:2010vs, Weinberg:2013kea, Kawasaki:2015ofa}). 
It is given as 
$g_* (T_D) \simeq 103.9$ for $T_D = 200 \GEV$, 
$g_* (T_D) \simeq 103.5$ for $T_D = 100 \GEV$, 
$g_* (T_D) \simeq 97.4$ for $T_D = 50 \GEV$, and 
$g_* (T_D) \simeq 86.2$ for $T_D = 10 \GEV$, 
which imply that 
the effective neutrino number is about 
$0.054 - 0.071$. 
The Planck data combined with the observation of BAO 
puts the constraint $N_{\rm eff} = 3.15 \pm 0.23$~\cite{Ade:2015xua}, 
which is consistent with the value predicted in the standard model ($N_{\rm eff} = 3.046$) 
and our prediction. 
The deviation from the standard value will 
be observed by the ground-based Stage-IV CMB polarization experiment 
CMB-S4, which measures $N_{\rm eff}$ with a precision of $\Delta N_{\rm eff} = 0.0156$ 
within $1 \sigma$ level~\cite{Wu:2014hta} (see also Ref.~\cite{Abazajian:2013oma}).

\subsection{Primordial abundance of charged particles}

When the reheating temperature of the Universe is higher than 
the freezeout temperature of $\psi$, its thermal relic abundance is determined as 
\beq
 \Omega_\psi h^2 \approx \frac{5.0 \times 10^{-27} {\rm \ cm}^3 {\rm s}^{-1}}{\la \sigma v \ra}, 
\eeq
where the annihilation cross section is given by 
\beq
 \la \sigma_\psi v \ra \simeq \frac{\pi \alpha_H^2 }{m_\psi^2} + N \frac{q_{\rm eff}^4 \pi \alpha_Y^2 }{m_\psi^2}. 
 \label{annihilation}
\eeq
The second term includes the annihilation into the SM particles 
and the prefactor $N$ is given by $N \simeq 1+ (5+1/8)/q_{\rm eff}^2$ 
for $m_\psi \gg \mathcal{O}(100 \GEV)$. 
Below we neglect the annihilation process via the EW force 
because its coupling constant is much smaller than 
that of hidden U(1)$_H$ ($\alpha_H \approx 0.1$). 
Note that the annihilation cross section increases by a factor of $1.2$ when 
we assume $q_{\rm eff} =2$ and take into account the annihilation into the SM particles. 
For typical parameters, 
their abundance is given by 
\beq
 \Omega_\psi h^2 \approx 0.013 \lmk \frac{m_\psi}{1 \  \TEV} \rmk^2 \lmk \frac{\alpha_H}{0.1} \rmk^{-2}. 
 \label{Omega_L}
\eeq

Fractionally charged particles may affect CMB temperature fluctuations, 
so that precise measurement of CMB temperature fluctuations 
provides an upper bound on their abundance. 
However, 
all of the previous works focused on the case of millicharged particles. 
Therefore their result cannot directly apply to our case, where the electric charge of $\psi$ 
is of order unity. 
Still, 
we expect that 
the abundance of these particles should be less than 
of order the uncertainty of baryon abundance determined by 
the Planck experiment. 
As discussed in Refs.~\cite{Dubovsky:2003yn, Dolgov:2013una}, 
their constraint comes from the fact that 
the exotic charged particles are tightly coupled with the plasma before the recombination epoch 
but the Compton scattering process is neglected due to its small electric charge. 
In particular, the constraint given in Ref.~\cite{Dolgov:2013una} is based mainly on the fact that 
increasing the number density of millicharged particles 
results in decreasing that of baryons, which results in decreasing that of electrons 
by the neutrality condition of the Universe. 
As a result, the Silk damping scale becomes larger compared with the case without millicharged particles. 
In our case, 
the tight coupling condition is satisfied due to the large electric charge 
while the Compton scattering process is neglected due to the suppression of the cross section 
by the large mass of $\psi$. 
Therefore we can apply their result to our case with $\mathcal{O}(1)$ electric charge. 
Thus we require~\cite{Dolgov:2013una}
\beq
 \Omega_\psi h^2 \lesssim 0.001. 
\eeq

Another constraint comes from the observation of Li abundance, 
which is marginally consistent with the prediction of the BBN theory 
without fractionally charged particles. 
When electrically charged particles are abundant in the BBN epoch, 
they form a bound state with $^4$He, which leads to 
an efficient production of $^6$Li via a photon-less thermal production process~\cite{Pospelov:2006sc, Kohri:2006cn}. 
The enhancement of $^6$Li production originates 
mainly from the fact that 
the Bohr radius of the bound stare is much shorter than 
the wavelength of emitted photon in the standard BBN theory. 
Since the Bohr radius is determined by the mass of nucleus 
and the charges of bounded particles, 
their results do not change by many orders of magnitude even in the case with 
fractionally charged particle with $O(1)$ electric charge. 
Thus we quote their results~\cite{Pospelov:2008ta}: 
\beq
 \frac{n_\psi}{n_b} \lesssim 10^{-5-6}, 
\eeq
where $n_b$ is the baryon number density. 
This constraint is severer than the one coming from the observations of CMB temperature fluctuations. 
We discuss how to evade these constraints in Sec.~\ref{possibilities}.

\subsection{Present abundance of charged particles}

Next, we consider an era after the solar system and the Earth form, 
following Ref.~\cite{Langacker:2011db}. 
The number density of fractionally charged particles in bulk matter 
(i.e., in the Earth or solar system) 
is different from that given in Eq.~(\ref{Omega_L}) because of their electrical interaction with matter, 
which results in efficient annihilation in bulk matter. 
However, 
the annihilation is not so efficient 
that we cannot avoid 
severe constraints by the null results of 
searches of fractionally charged particles in bulk matter.

Fractionally charged particles in the Earth 
are more dense than their average density in the Universe 
because they behave like baryons due to their electric charge. 
In addition, the annihilation cross section is enhanced by the Sommerfeld enhancement 
effect in a low terrestrial temperature: 
\beq
 \la \sigma_\psi v \ra_{\rm SF} = S \lmk \eta \rmk \la \sigma_\psi v \ra, 
\eeq
where $S(\eta)$ is a Sommerfeld enhancement factor 
given by 
\beq
 S(\eta) \equiv \frac{\eta}{1 - e^{-\eta}}. 
\eeq
The parameter $\eta$ is defined by 
\beq
 \eta \equiv 2 \pi \frac{\alpha_H}{\beta}, 
\eeq
where $\beta$ is the velocity of the fractionally charged particle 
in a low terrestrial temperature of order $300$ K. 
Since the time scale is the age of the Earth, which is of order 
$4.5$ Gyr ($\equiv t_{\rm E}$), 
the annihilation reduces the abundance of fractionally charged particles 
to the amount of 
\beq
 \lmk \frac{n_\psi}{n_B} \rmk \simeq \frac{1}{n_B \la \sigma_\psi v \ra_{\rm SF} t_{\rm E}}. 
\eeq
Using the number density of baryons in bulk matter of $n_B \simeq 6.4 \times 10^{23} {\rm \ cm}^{-3}$, 
we obtain 
\beq
 \lmk \frac{n_\psi}{n_B} \rmk \simeq 8.5 \times 10^{-24} 
 \lmk \frac{m_\psi}{1 \ \TEV} \rmk^{3/2} 
 \lmk \frac{\alpha_H}{0.1} \rmk^{-3}. 
 \label{n_L per baryon}
\eeq

One may wonder that 
negatively fractionally charged particles capture 
protons and/or Heliums and form 
positive exotic ions, 
which cannot annihilate with anti-particles due to the electrical repulsion of 
Coulomb force~\cite{Langacker:2011db}. 
However, 
in our model, 
the annihilation occurs due to the hidden U(1)$_H$ gauge interaction, 
which is much stronger than the electric force, 
so that 
annihilation cannot be prevented by the visible Coulomb force. 
Therefore the abundance of fractionally charged particles in the Earth 
is given by Eq.~(\ref{n_L per baryon}).

Fractionally charged particles can be observed by searching in bulk matter 
if they are trapped in rigid matter or water. 
Most of the searches of fractionally charged particles put constraints on the abundance of 
particles with a fractional charge in the interval of $[n + 0.2, n + 0.8]$ 
where $n$ is any integer~\cite{Lee:2002sa, Kim:2007zzs} (see Ref.~\cite{Perl:2009zz} for a review). 
In the recent paper of Ref.~\cite{Moore:2014yba}, however, 
they provided a constraint which is less stringent but is applicable to broader range of charges 
by using optically levitated microspheres in high vacuum. 
They also claimed that the previous works can constrain 
the abundance of particles with smaller charges 
by assuming the abundance of negative fractionally charged particles. 
This is because 
there can be multiple fractionally charged particles in each sample 
when their number density is sufficiently large. 
As a result, the total charge in each sample is the summation of 
charges of those particles, which can be larger than about $0.2$ 
and can be detectable by those experiments. 
Their results indicate that 
the abundance of fractionally charged particles has to be 
$15$\,-\,$23$ order of magnitude less than that of baryons, depending on their charge. 
For example, 
for a fractional charge in the interval of $[n + 0.2, n + 0.8]$ 
abundance per nucleon should be less than $10^{-23}$, 
and 
for a fractional charge in the interval of $[n + 0.1, n + 0.2]$ and $[n + 0.8, n + 0.9]$
abundance per nucleon should be less than $10^{-20-21}$. 
The result of Eq.~(\ref{n_L per baryon}) is consistent with this upper bound. 
However, 
we should also consider the abundance of fractionally charged particles 
in the interstellar medium (ISM). 
Its number density in the ISM is much less than that in the Earth, 
so that the annihilation is inefficient as discussed in Ref.~\cite{Langacker:2011db}. 
This may imply that 
the searches of fractionally charged particles in meteorites (e.g., the work of Ref.~\cite{Kim:2007zzs}) 
exclude our model 
though the evolutionary history of meteoritic material is uncertain. 
In addition, 
the calculation of Eq.~(\ref{n_L per baryon}) does not take into account 
the flux of fractionally charged particles from the outer region of the Earth. 
Since they are abundant in the outer region, 
fractionally charged particles may fall into the Earth just like cosmic rays. 
Therefore 
the constraints coming from the search in bulk matter may exclude 
the scenario that fractionally charged particles survive at present. 
In the next subsection, we provide some mechanisms to evade those constraints.

\subsection{Possibilities to eliminate unwanted relics \label{possibilities}}

We can consider a scenario in which the unwanted charged particles 
are never produced after inflation, 
which requires that the maximal temperature of the Universe after inflation 
is much lower than the mass of $\psi$.%
\footnote{
In Ref.~\cite{Mukaida:2015ria}, 
one of the authors (M.Y.) investigated the thermalization process of inflaton decay products 
and found that the maximal temperature of the Universe after inflation 
can be much lower than the one expected in the literature due to the delay of thermalization. 
It was found that the maximal temperature can be less than $100 \ \GEV$. 
} Alternatively, one may consider huge late-time entropy production by thermal inflation.
In this case, the hidden photon is also diluted, 
so that the dark radiation is absent [see Eq.~(\ref{delta N_eff})].

Another way to evade the constraints is to enhance their annihilation rate by 
a strong interaction. 
We may introduce an additional U(1)$_{H2}$ gauge symmetry 
under which the fields $\psi_i$ are charged~\cite{Yamada:2015waa,Kawasaki:2015lpf}. 
We also introduce a scalar monopole that develops an expectation value of order $1  \ \TEV$ 
to break U(1)$_{H2}$ spontaneously.\footnote{
Since $\psi_i$ has both charges of $U(1)_H$ and $U(1)_{H2}$, an operator 
$g^{\rho \sigma} |\phi_M|^2  {F_{\mu \nu}^H} F_{\rho \sigma}^H$ arises where $\phi_M$ is a scalar monopole.
Therefore, it may be possible to test a TeV scale photon-photon collider depending on the sizes of the operator and $\chi$.
} 
As a result, 
the fields $\psi_i$ are connected by the physical string due to the dual Meisner effect~\cite{Nambu:1974zg}, 
so that they soon annihilate with each other after the spontaneous symmetry breaking (SSB). 
Therefore the fields $\psi_i$ are absent after the SSB, 
so that we can evade the constraint coming from the searches of bulk matter 
as well as that coming from the observations of CMB temperature fluctuations.%
\footnote{
The U(1)$_{H2}$ gauge theory may be conformal 
in the presence of monopole as well as electrons. 
Thus we need to take care of anomalous dimension of U(1)$_{H2}$ gauge field strength, 
which may result in the absence of interactions between U(1)$_{H2}$ and U(1)$_H$~\cite{Argyres:1995xn}. 
In this paper 
we neglect this issue 
by assuming that 
the conformal coupling constant is not large and the anomalous dimension is sufficiently small. \label{ft:yamada}
}
We assume that 
there is no additional mixing among U(1)$_Y$ and U(1)$_{H2}$
so that our calculations given in the previous sections are not changed.\footnote{
A kinetic mixing between U(1)$_H$ and U(1)$_{H2}$ can be removed by the shift of U(1)$_{H2}$ charge of $\psi_i$. 
}
In this case, when $\psi$ and $\bar \psi$ are produced via the Drell-Yan process at the collider experiment,
they form a bound state decaying into $\gamma \gamma$, $\gamma \gamma'$ and $\gamma' \gamma'$.
Also, when the center of energy available  is sufficiently large,
mesonization occurs and four-photon signals will be observed 
due to the subsequent annihilation of the mesons $(\psi \bar\psi)$.

When the number of flavor of the hidden particle is larger than unity, 
we can predict a long-lived neutral particle 
that can be a candidate for DM. 
We consider $\psi_i$ with $i=1,2$ and assume that the flavour is not mixed 
so that the $\bar \psi_1 \psi_2$ bound state (which we denote $\pi_{\rm DM}$) 
does not annihilate after the SSB of U(1)$_{H2}$. 
Since this bound state is neutral and stable, 
it can be DM. 
When their masses are larger than $v$, 
the relic abundance of the fields $\psi_i$ is determined by their annihilation rate [see Eq.~(\ref{Omega_L})]. 
Then they are attached by the physical string after the SSB. 
The other bound states (e.g., $\bar{\psi}_1 \psi_1$) annihilate into visible photons. 
As a result, the relic density of $\psi$ is given by 
\beq
 n_\psi \sim \frac{n_{\psi_1} n_{\psi_2}}{n_{\psi_1} + n_{\psi_2}}. 
\eeq
This is consistent with the observed DM abundance 
when the masses of $\psi_i$ are of order $1 \TEV$.%
\footnote{
On the other hand, 
when their masses are smaller than $v$, 
bound states form at the SSB and then their abundance is determined by the 
subsequent annihilation. 
The annihilation rate of $\pi_{\rm DM}$ is estimated as $\Lambda^{-2}$ 
where $\Lambda$ ($\approx 4 \pi v$) is the dynamical scale of the confinement. 
Thus we can account for the observed DM density 
when the dynamical scale is around the unitarity bound of order $100 \  \TEV$. 
In this case, however, 
the cross section of the diphoton signal is suppressed 
and we cannot explain the excess reported by ATLAS and CMS. 
}
The DM can decay when 
we write a higher dimensional operator of $\psi_1 \bar{\psi}_1 \psi_1 \bar{\psi}_2 / M_{\rm pl}^2$. 
However, 
its lifetime is much larger than the present age of the Universe for $\Lambda \lesssim 100 \ \GEV$, 
so that we expect no astrophysical signal from DM decay.

\section{Discussion and conclusion}
\setcounter{equation}{0}

We have shown that the $750$\,GeV diphoton excess can be explained by introducing a pair of vector-like
quarks and hidden fermions charged under a hidden U(1) gauge symmetry which has a significantly mixing with U(1)$_Y$.
The hidden U(1) has a relatively large coupling constant, which may naturally arise from string theory compactifications~\cite{Tatar:2008zj}. Due to the large coupling and kinetic mixing, hidden fermion loops induce a sizable branching fraction of a 750\,GeV scalar to diphotons. Notably, the standard model gauge couplings unify around $10^{17}$\,GeV with effects of the kinetic mixing, suggesting the grand unified theory without too rapid proton decay. 
Obviously, instead of introducing the hidden fermions, one can consider vector-like quarks (and leptons) charged under both the SM gauge symmetry and the hidden U(1) gauge symmetry, which leads to similar results.

Our scenario can be checked by looking for events with a photon and missing transverse momentum.
Its cross section is related to that for the diphoton final state through the kinetic mixing, 
and predicted to be large.  Therefore, our scenario can be tested in the near future at the LHC experiment.

So far we have focused on the case in which the phase component, $\phi$, is responsible for
the diphoton excess. As mentioned earlier, our analysis can be similarly applied to the radial
component, $s$, in Eq.\,(2.9); in general, $s$ with a mass of 750 GeV decays mainly into a pair of $\phi$,
each of which decays into $\gamma \gamma$, $\gamma \gamma'$, and $\gamma' \gamma'$,
in addition to the loop-induced decays. 
The photons ($\gamma\gamma$) produced from the decay of $\phi$ are
collimated if the axion is sufficiently boosted, and the two collimated photon pairs may be
identified with the diphoton signal in the detector analysis. 
(See Refs.~\cite{Knapen:2015dap,Agrawal:2015dbf,
Chala:2015cev,Aparicio:2016iwr,Ellwanger:2016qax,Dasgupta:2016wxw,Chiang:2016eav}
 for the collimated photons in association with the diphoton excess.)  
 In our model, the ratio of the branching fraction into the combination of $\gamma$ and $\gamma'$ depends on the kinetic mixing. 
As a result, we may have triphoton and tetraphoton signals, depending on the probability that the collimated photons are identified with a single photon at the detector.
If those signals are confirmed by the upcoming LHC data, it would give a smoking gun
signature of the diaxions decaying into photons and hidden photons.
Further analysis of the above processes is warranted.

\section*{Acknowledgments}
M. Y. thanks T. T. Yanagida and K. Yonekura for useful comments concerning footnote~\ref{ft:yamada}.
This work is supported by MEXT KAKENHI Grant Numbers 15H05889 and 15K21733 (F.T. and N.Y.),
JSPS KAKENHI Grant Numbers  26247042, and 26287039 (F.T.), 
JSPS Research Fellowships for Young Scientists(M.Y.), 
World Premier International Research Center Initiative (WPI Initiative), MEXT, Japan (F.T. and M.Y.).



\end{document}